# "Quantum" Chaos and Stability Condition of Soliton-like Waves of Nuclear Burning in Neutron-Multiplicating Media

V.D. Rusov[1,2], E.P. Linnik[1], V.A. Tarasov[1], T.N. Zelentsova[1],
I.V. Sharf[1], S.A. Chernezhenko[1], O.A. Byegunova[1,2],

[1]*Odessa National Polytechnic University, 65044, Odessa, Ukraine*
[2]*Bielefeld University, D-33615, Bielefeld, Germany*

## Abstract

We show that the stability condition for the soliton-like wave of nuclear burning in neutron-multiplicating medium is determined in general by two conditions. The first condition (necessary) is determined by relationship between the equilibrium concentration and critical concentration of active (fissile) isotope, that is a consequence of the Bohr-Sommerfeld quantization condition. The second condition (sufficient) is set by the so-called Wigner quantum statistics, or more accurately, by a ststistics of the Gaussian simplectic ensembles with respect to the parameter that describes the squared width of burning wave front of nuclear fuel active component.

______________________________________________________________

[*] Corresponding author: Rusov V.D., E-mail: siiis@te.net.ua

# I. INTRODUCTION

In spite of obvious efficiency and allurement of the nuclear power engineering of next generation, the main difficulties of its perception are predetermined by non-trivial properties which future ideal nuclear reactor must possess. At first, the natural, i.e. unenriched uranium or thorium must be used as a nuclear fuel. Secondly, the reactivity regulation system of reactor by traditional control rods is completely absents, but for all that a reactor must possess the property of so-called inner safety. It means that the critical state of reactor core must be permanently maintained in any situation, i.e. the reactor normal operation is automatically maintained not as a result of operator activity, but by virtue of physical reasons-laws preventing the explosive development of chain reaction by the natural way. Figuratively speaking, the reactor with inner safety it is "the nuclear installation which never explode" [1].

$$^{238}U(n,\gamma) \to {}^{239}U \xrightarrow{\beta^-} {}^{239}Np \xrightarrow{\beta^-} {}^{239}Pu(n, fission) \qquad (1)$$

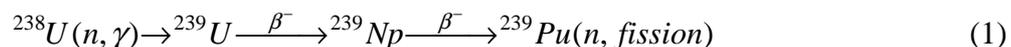

Strangely enough, but reactors satisfying such unusual requirements are possible in the reality. For the first time the idea of such reactor was proposed by Feoktistov [2] and independently by Teller, Ishikawa and Wood [3].

The main idea of reactor with inner safety consists in the selection of fuel composition so that, at first, the characteristic time $\tau_\beta$ of the nuclear burning of fuel active (fissile) component is substantially greater than the characteristic time of delayed neutrons production and, secondly, necessary self-regulation conditions are meet during the reactor operation (that always take place, when the equilibrium concentration $\tilde{n}_{fis}$ of fuel active component is greater than critical concentration $n_{crit}$ [2]). These very important conditions can practically always to be attained, if among other reactions in the reactor the chain of nuclear transformations of the Feoktistov uranium-plutonium cycle type [2]

$$^{238}U(n,\gamma) \to {}^{239}U \xrightarrow{\beta^-} {}^{239}Np \xrightarrow{\beta^-} {}^{239}Pu(n, fission) \qquad (1)$$

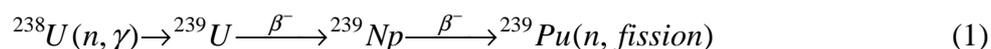

or the Teller-Ishikawa-Wood thorium-uranium cycle type [3]

$$^{232}Th(n,\gamma) \to {}^{233}Pa \xrightarrow{\beta^-} {}^{233}U(n, fission), \qquad (2)$$

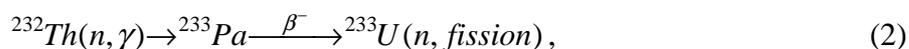

will be enough appreciable.

In both cases the produced fissile isotopes of $^{239}$Pu or $^{233}$U are the active components of nuclear fuel. The characteristic time of such reaction, i.e. the time of proper $\beta$-decays, is approximately equal to $\tau_\beta = 2.3/\ln 2 \approx 3.3$ days for reaction (1) and $\tau_\beta \approx 39.5$ days and for reaction (2), that is several orders greater than the time of delayed neutrons production.

The self-regulation of nuclear burning process is stipulated by the fact that such system left by itself can not pass from a critical state to reactor acceleration mode, because a critical concentration is bounded from above by the finite equilibrium concentration of nuclear fuel fissile component (plutonium for (1) or uranium for (2)), i.e. $\tilde{n}_{fis} > n_{crit}$ (Feoktistov's stability condition [2]). On phenomenological level the self-regulation of nuclear burning is manifested as follows. The increase of neutron flux due to some reasons will result in the rapid burnup of nuclear fuel fissile component (plutonium for (1) or uranium for (2)), i.e. its concentration as well as the neutron flux will decrease, while the new nuclei of corresponding fissile component of nuclear fuel are produced with the same generation rate during time $\tau_\beta$. And vice versa, if the neutron flux is sharply decreased due to external action, the burnup rate decrease too, and the accumulation rate of fuel fissile component will be increased as well as the number of neutron production after a while $\tau_\beta$.

However, as is known [2], the Feoktistov stability condition is only necessary but insufficient condition. Therefore full generalization of the Feoktistov stability condition for critical waves of nuclear burning in neutron-multiplicating mediums is the purpose of this paper.

## II. PROPERTIES OF STABILITY CONDITION FOR CRITICAL WAVE OF NUCLEAR BURNING ACCORDING TO FEOKTISTOV

Following [2], let us consider the known "polygon" system of kinetic equations for neutrons and nuclei in the reaction chain (1) with respect to the normalized autowave variable $z=(x+ut)/L$:

$$\frac{d^2 n_*}{dz^2} = \left[1 - \frac{n_{Pu}}{n_{crit}^{Pu}}\right] n_*, \qquad (3)$$

$$\Lambda \frac{dn_8}{dz} = -[n_8 - n_9 - n_{Pu}] n_*, \qquad (4)$$

$$\Lambda \frac{dn_9}{dz} = (n_8 - n_9) n_* - n_9, \qquad (5)$$

$$\Lambda \frac{dn_{Pu}}{dz} = n_9 - \frac{n_{Pu}}{\tilde{n}_{Pu}} n_*, \qquad (6)$$

where $u$ is phase velocity of the steady running wave, $L$ is the neutron average diffusion length, $n_*(z,t)$ is the neutron density, $D = v/3\Sigma_s = L^2/\tau$ is neutron diffusion constant, $cm^2 \cdot s^{-1}$; $v$ is neutron velocity in the one-group approximation, $cm \cdot s^{-1}$; $\Sigma_s$ is neutron microscopic scattering cross-section, $cm^{-1}$; $\tau = 1/v\sum \sigma_a^i N_i$ is neutron lifetime in medium, $s$; $\Lambda = u\tau_\beta/L$ is dimensionless constant, $n_{crit}^{Pu} = N_{crit}/N_8(-\infty) = \sum \sigma_a^i n_i)/(\nu-1)\sigma_f^{Pu}$ is the plutonium relative critical concentration, $N_{crit}$ is the plutonium critical concentration, $N_8$ is the $U^{238}$ concentration, $\sigma_a$ and $\sigma_f$ are the microscopic neutron capture cross-section and fission cross-section, respectively, $n_8$ and $n_9$ are the concentrations of $U^{238}$ and $U^{239}$ normalized to $U^{238}$ initial concentration, i.e., to $N_8(-\infty)$, $\nu$ is the average number of prompt neutrons produced per plutonium nucleus fission.

Solving these equations Feoktistov was based on the analogy of diffusion equation and the Schrödinger steady-state equation in quasi-classical approximation [2]. Naturally, in this case (see Eq. (3)) the stationarity condition of solution is satisfied integrally, because there are points where $n_{Pu} > n_{crit}$, and there are points where $n_{Pu} < n_{crit}$. In this sense, the region at $n_{Pu} > n_{crit}$ corresponds as it were to allowed region, while the region at $n_{Pu} < n_{crit}$ corresponds to subbarrier region. In other words, the inverted profile of plutonium concentration in the $^{238}U$ medium plays the role of potential well (Fig. 1(a) [4]).

In the region at front of wave ($z = -\infty$) the approximate solution looks like

$$n = C \exp z, \tag{7}$$

$$n_8 = \exp\left(-\frac{C}{\Lambda}\exp z\right), \tag{8}$$

$$n_9 = \frac{C}{1+\Lambda}\exp z, \tag{9}$$

$$n_{Pu} = \frac{\tilde{n}_{Pu}}{1+\Lambda}\left[1 - \exp\left(-\frac{C}{\Lambda \tilde{n}_{Pu}}\exp z\right)\right]. \tag{10}$$

Let us remind that obtaining this solution, we have neglected summands $n_9$ and $n_{Pu}$ whose values are determined by edge condition $n_8 \cong 1$. Then assuming that the subbarrier region ends at $z=0$, we have $n_{Pu} = n_{crit}$ at this point. This allows us to determinate the value of constant C. At the point $z=a$, according to the Bohr-Sommerfeld quantization condition, we have the following equality

$$\int_0^a \sqrt{\frac{n_{Pu}}{n_{crit}^{Pu}} - 1}\, dz = \frac{\pi}{2}, \tag{11}$$

where the integral is taken over the supercritical region ($n_{Pu} > n_{crit}$). At the same time condition (11) plays also the role of condition for finding the point $a$ at $n_{Pu} = n_{crit}$, i.e., when the transition into subbarrier region happens due to burn-up (see Fig. 1(a) and Fig.2)[1].

Executing the ordinary for quasi-classical approximation linkage with the supercriticality region ($n_{Pu} > n_{crit}$) we will come to calculation of $\Lambda$.

As a critical state is automatically maintained at $n_{Pu} > n_{crit}$ [2] (that is the direct consequence of the Bohr-Sommerfeld quantization condition), we can use this fact for generalization of the following inequality:

$$\tilde{n}_{Pu} > n_{Pu} > n_{crit}^{Pu}, \qquad (12)$$

Thus, Feoktistov shown for the first time [2] that the soliton-like propagation of neutron-fission wave of nuclear burning is possible in $^{238}$U medium only under the condition of a certain ratio between equilibrium and critical plutonium concentrations ($\tilde{n}_{Pu} > n_{crit}$), which is characterized by the Bohr-Sommerfeld quantization condition. In other words, only in this case the critical (quasi-stationary) state of system (reactor core) can automatically maintained without any external intervention, and, consequently, only in this case the reactor fully and unambiguously possesses the inner safety properties.

It is appropriate here to pay an attention to very important Feoktistov's parameter, which, as shown below, is basis for ideology of the stability of soliton-like wave of nuclear burning:

$$\Lambda(a) = \frac{u\tau_\beta}{L}, \qquad (13)$$

where $a$ is the width of permitted range of integration in the Bohr-Sommerfeld condition (11), where the inequality $n_{Pu} > n_{crit}$ (Fig. 2) and $\tilde{n}_{Pu} > n_{crit}$, respectively, are satisfied; $\Lambda(a)$ is dimensionless coefficient, which appears within the framework of simplified diffusion model of the Feoktistov reactor (3)-(6).

---

[1] Note that the model calculations of the Feoktistov problem by the system of equations (3)-(6) really show [4] that at steady-state conditions the Bohr-Sommerfeld quantization condition is fulfiled with an accuracy up to a few percents (!!!). Authors [4] note that there are no grounds to expect the more exact coincidence because a quantization condition for lower level is approximate.

Obviously, Eq. (11) due to its physical meaning is a key factor which predetermines the phase velocity of soliton-like burning wave. Therefore, this equation exists regardless of an idealization degree of reactor core model and should appear in explicit or implicit form in any model whose the system of kinetics equations for neutrons and nuclei has soliton-like solutions for neutrons. At the same time, as the average width of soliton wave has an order of $2L$, the maximum values of the dimensionless coefficient $\Lambda(a)$ and wave velocity $u$ are determined by the following approximate equality

$$\frac{1}{b}\Lambda_{max}(a) = \frac{u_{max}\tau_\beta}{bL} = 1, \qquad (14)$$

where coefficient is $b \sim 2$ although a final estimation will be done below.

From analysis of Eq. (14) it follows that the velocity of stable propagation of soliton-like wave is not necessarily equal to the diffusion rate $u=L/\tau_\beta$. It can be considerably slower or faster due to very strong domination either of the nonlinearity parameter or dispersion parameter, which in its turn reflects the peculiarities of nuclear transformation kinetics, for example, in the chain (1) and/or in (2). In practice they manifest itself as higher or lower degree of fuel burn-up.

In other words, when the wave velocity and consequently the degree of fuel burnup are low, the wave stops due to the following reasons. Neutrons from an external source, which take place in the initial stage of wave initiation, burn out the plutonium on medium bondary and simultaneously transmute the uranium into $^{239}$Np. Neptunium with time starts to produce the plutonium but it can not create the required high concentration, while the $^{239}$Pu production decreases due to the uranium burnup. More and more thick layer without both $^{238}$U and $^{239}$Pu grows on the medium boundary. The neutron diffusion through this layer does not provide the increase of plutonium concentration in next layers, and the wave does not arise even at $n_{Pu}(x,0)= n_{crit}$.

Conversely, when the wave velocity and degree of fuel burn-up are high, the wave stops also because of the scarce (or more exactly, delayed) plutonium production which takes place due to another reason. Figuratively speaking, the situation resembles the fire in the forest under strong wind, when only tree crowns burn. When the wind speed increases, it could extinguish the fire at all. We have the similar situation, when there is a velocity, at which in the early stage (when $x \approx 0$) the front of neutron soliton wave outruns the front of plutonium production wave, and this advance exceeds the neutron diffusion length. This leads, in fact, to transformation of fast wave into slow wave or to its full stop. It is interesting to note that this case not studied in the literature (with the exception of [4,5]), but it is possible to postulate that it corresponds to some hypothetical situation, when the nuclear burning wave forms in highly-enriched fuel which has the ultra-low critical concentration of fuel fissile component.

Thus, the lag (Fig. 1(b)) or advance of neutron wave front relative to the plutonium wave front for a distance considerably exceeding the neutron diffusion length will leads to stop and total degradation of these waves. This means that degradation of waves with very low or very high initial phase velocity will exhibits as the tendency to zero of Eq. (11) at very low or very high values of *a*. Therefore taking into account Eq. (14), we can conclude that Eq. (11) is true in the range $0 \leq (1/b)\Lambda(a) \leq 1$. Based on this generalization, we can make an important assumption that the expression $(1/b)\Lambda(a)$ means the certain probability density distribution $p(a)$ with respect to *a*:

$$\frac{u\tau_\beta}{bL} = p(a). \tag{15}$$

Let us consider and substantiate the type and main properties of such a statistics, and also show the results of its verification based on the known computational experiments on simulation of nuclear burning wave in the U−Pu (1) and Th−U (2) fuel cycles.

## III. CHAOS AND INTEGRABILITY IN NONLINEAR DYNAMIC OF REACTOR CORE

In order to solve the assigned task we use the known analogy between the neutron diffusion equation and the Schrödinger steady-state equation in quasiclassical approximation. We would remind that this analogy was used earlier to solve the system of kinetics equation for neutrons and nuclei (3)-(6) in the reaction chain (1) of the U−Pu fuel cycle. Since the system of equations for neutrons and nuclei in the Th−U fuel cycle (2) is structurally identical to the system equation for the U−Pu fuel cycle (1), the computed "quantum mechanical" solution, which describes the statistics (15), will be general for both fuel cycles, except for a few details.

Now, let us remind that earlier we have used the Bohr-Sommerfeld quantization condition which in the case of the one-dimensional systems determines in the explicit form the energy eigenvalues $E_n$

$$\oint p(x)dx = \oint \sqrt{2m(E_n - V(x))}dx = 2\pi\hbar\left(n + \frac{1}{2}\right), \quad n = 0, 1, 2..., \tag{16}$$

where *m* and *p(x)* are the mass and momentum of particle in the field of some smooth potential *V(x)*.

For the Feoktistov nearly integrable system of the equations (3)-(6) or for the anologous Teller system of equations, for which it is assumed that $m=1/2$, $V(x)=1$ and $n=0$, this condition is applied in the form

$$\int_0^a \sqrt{E_0 - 1} dz = \frac{\pi}{2}, \quad E_0 = \frac{n_{fis}}{n_{crit}^{fis}}, \quad (17)$$

where index *fis* denotes the fissionable isotope, for example, the $^{239}$Pu in the Feoktistov U−Pu fuel cycle (1) or the $^{233}$U in the Teller Th−U fuel cycle.

However, in describing the real evolution of fast reactor core, the corresponding systems of equations for neutrons and nuclei are nonintegrable almost without exception. This, in its turn, means that according to the Kolmogorov-Arnold-Moser theorem [6,7] quasiclassical quantization formulas are inapplicable for the system, where the motion in phase space is not limited by multidimentional tori. This is stipulated by the fact that in the Hamiltonian nonintegrable systems the more and more number of tori collapse in phase space with perturbation (nonintegrability) growth. As a result, the trajectories of majority of bound states gets entangled, the motion becomes mainly chaotic, and bound states themselves and their energies, can not be described by the rules of quasiclassical quiantization, for example, such as the Einstein-Brillouin-Keller (EBK) quantization rule for multidimentional case [7,8], which generalizes the Bohr-Sommerfeld quiantization rule. Note that nowadays a notion "quantum chaos" is included the circle of problems related to quantum-mechanical description of systems chaotic in a classic limit [9, 10].

Since the results of random matrices theory will be used for research of chaotic properties of the statistics (11), we first give an overview of the main concepts of this theory.

First, following [9,10], let us shortly consider a nature of so-called universality classes and the Gaussian ensemble types. As is known, the Hamilton operator matrix in possession of any kind of a symmetry can be reduced to the block-diagonal form. At the same time, matrix elements in each block are specified by a certain quantum number set. For the sake of simplicity we assume that the Schrödinger equation $i\hbar(\partial \psi / \partial t) = \hat{H}\psi$ is expressed for states belonging to the one block. At the same time the size of the operator $\hat{H}$ matrix is finite and equal to an integer.

As shown in [9,10], these universality classes separate physical systems into groups according to their relation to orthogonal, unitary or simplectic transformation, which leave the $\hat{H}$ matrix invariant. In other words, as it postulated in [9]:

• the Hamiltonian of spinless system possessing a symmetry with respect to time inversion is invariant under orthogonal transformations and can be represented by real matrix;

• the Hamiltonian of spinless system not possessing a symmetry with respect to time inversion is invariant under unitary transformations and can be represented by the Hermitian matrix;

• the Hamiltonian of the system with spin of 1/2 possessing a symmetry with respect to time inversion is invariant under simplectic transformations and can be represented by quaternion real matrix.

Now let us talk about the Gaussian ensembles. If the matrix element distribution function is invariant under one of indicated transformations, this means that the sets of all matrices with elements described by these distribution functions form the Gaussian orthogonal ensemble (GOE), the Gaussian unitary ensemble (GUE) and the Gaussian simplectic ensemble (GSE), respectively.

At the same time it should be noted the one very substantial detail. The matrix element distribution function of the Gaussian ensembles can not be directly measured, since the experiment can give us information about the energy levels of investigated quantum-mechanical system only. In other words, just the energy eigenvalues distribution function is of greater interest from the practical point of view.

Derivation of corresponding equations for the considered types of the Gaussian ensembles can be found in [10]. At the same time, the correlated distribution function of energy eigenvalues it is possible to write down in the sufficiently universal form for all ensemble types:

$$P(E_1,...,E_N) \sim \prod_{n>m}(E_n - E_m)^\nu \exp(-A\sum_n E_n^2), \qquad (18)$$

where $\nu$ is an universality index, which takes on the value of 1, 2 and 4 for GOE, GUE and GSE statistics, respectively. At $\nu=0$ energy eigenvalues are not correlated. In this case, the energy level spacing distribution function is described by the Poisson statistics, and the matrix ensemble itself is called the Poisson ensemble.

So long as the energy level spacing distribution function is the most studied property of chaotic systems, following [9], we give a calculation only for relatively simple case of the Gaussian ensemble with matrixes 2×2 in size. Let us calculate the energy level spacing distribution function $p_W(s)$ substituting the function $P(E_1, E_2)$ in (18):

$$p_W(s) = \int_{-\infty}^{+\infty}dE_1 \int_{-\infty}^{+\infty}dE_2 P(E_1,E_2)\delta(s-|E_1 - E_2|) =$$

$$= C \int_{-\infty}^{+\infty} dE_1 \int_{-\infty}^{+\infty} dE_2 |E_1 - E_2|^\nu \exp(-A \sum_n E_n^2) \delta(s - |E_1 - E_2|). \quad (19)$$

Constants *A* and *C* are defined by the two normalization conditions:

$$\int_0^\infty p_W(s) ds = 1, \quad (20)$$

$$\int_0^\infty s p_W(s) ds = 1. \quad (21)$$

The first condition is normalization of the total probability, and the second condition is normalization of the average energy level spacing. Integration of (19) gives us the so-called Wigner energy level spacing distribution functions, which correspond to the different Gaussian ensembles:

$$p_W(s) = \begin{cases} \dfrac{\pi}{2} s \exp(-\dfrac{\pi}{4} s^2), & \nu = 1 (GOE); \\ \dfrac{32}{\pi} s^2 \exp(-\dfrac{\pi}{4} s^2), & \nu = 2 (GUE); \\ \left(\dfrac{8}{3\sqrt{\pi}}\right)^6 s^4 \exp(-\dfrac{64}{9\pi} s^2), & \nu = 4 (GSE). \end{cases} \quad (22)$$

Despite the fact that these functions were obtained for the Gaussian ensemble with matrixes 2×2 in size, they describe with sufficient accuracy the spectra of arbitrary size matrices [9].

Note that random matrix theory at first was developed to find some regularities of heavy nucleus energy spectra [10,11], but it attracted keen interest after the Bohigas, Giannoni and Schmit conclusion [12] that this theory can be applied to any chaotic system.

We now turn to our problem of determination of statistics (15) type and will try to use the considered statistics properties of the Gaussian ensembles.

## IV. THE WIGNER QUANTUM STATISTICS AND GENERALIZED STABILITY CONDITION

Now, in the framework of nearly integrable system, to which the system of equations describing the nuclear burning kinetics of the Feoktistov U−Pu fuel cycle (1) or the Taylor Th−U fuel cycle (2) belongs, we formally introduce the "energy" eigenvalue of stationary state as $(\tilde{n}_{fis0}/n_{crit}^{fis})=E_0$ and "energy" eigenvalue of quasistationary state as $(n_{fiss}^{semi}/n_{crit}^{fis})=E_{semi}$ (where $E_0 \geq E_{semi}$ and $\tilde{n}_{fis0}$ is the current equilibrium concentration of fissile isotope limited from above by its initial equilibrium concentration, i.e., $\tilde{n}_{fis0} < \tilde{n}_{fiss}$). In general case, to describe the wave mode of nuclear burning, when the reactor is maintained in the near-critical state, we can consider that $E_{semi} \rightarrow 1$. Then in the framework of quantum-mechanical analogy, this means that the evolution of nuclear burning "energy" spectrum in allowed region is described by some quasi-equivalent two-level scheme (Fig. 3).

Then, for the nearly-integrable system which describes the nuclear transformation kinetics for the Feoktistov (1) or for the Teller (2) fuel cycle in general case we can use the Bohr-Sommerfeld approximate condition in the form

$$\int_0^a \sqrt{\frac{n_{fis}}{n_{crit}^{fis}} - 1} \, dz \approx a\sqrt{E_0 - E_{semi}} \sim \frac{\pi}{2}. \tag{23}$$

It follows that, we can postulate one obvious and important assertion: by virtue of the Bohr-Sommerfeld condition (23) the type of the Wigner energy level spacing statistics unambiguously predetermines the analogous statistics type of parameter, which characterizes the squared width ($a^2$) of concentration wave front of active (fissile) material.

Note that we have not any information about the value of energy $E_0$ before the experiment, whereas it is possible to consider that $E_{semi} = 1$. If to add also, that in the steady-state mode all wave kinetic parameters are predetermined by the initial equilibrium $\tilde{n}_{fis}$ and critical $n_{crit}^{fis}$ concentration of active (fissile) isotope (whose values are known before experiment), the physical meaning and the necessity of following change

$$a\sqrt{E_0 - E_{semi}} = a_* \sqrt{\frac{\tilde{n}_{fis}}{n_{crit}^{fis}} - 1} \tag{24}$$

become apparent.

It is obvious that the conditions (23) and (24) make it possible to obtain the expression for parameter $a_*$:

$$a_*^2 \sim \frac{\pi^2}{4} \frac{n_{crit}}{\tilde{n}_{Pu} - n_{crit}}. \tag{25}$$

The next step for determining the statistics $p(a_*)$ of Eq. (15) type consists in the experimental validation of proposed hypothesis. For that we have compared the Gaussian ensemble statistics (22) with the calculated data of well-known computational experiments [4,13-17] and have obtained a good accordance of calculation data with theoretical dependence, which is described by the Gaussian simplectic ensemble statistics (see Table I and Fig. 4).

Thus, we can conclude that the wave velocity (15) is predetermined by the following approximate equality

$$\frac{u\tau_\beta}{2L} \cong p_W^s(a_*) = \left(\frac{8}{3\sqrt{\pi}}\right)^6 a_*^4 \exp\left(-\frac{64}{9\pi}a_*^2\right), \quad a_*^2 \cong \frac{\pi^2}{4} \cdot \frac{n_{crit}^{Pu}}{\tilde{n}_{Pu} - n_{crit}^{Pu}}, \tag{26}$$

where coefficient $b = 2$ (see Eq.(15)); $\tau_\beta$ is the delay time caused by active (fissile) isotope production, which is equal to the $\beta$–decay time of compound nuclei in the Feoktistov (1) or the Teller (2) fuel cycle; $p_W^S(a_*)$ is the Wigner symplectic statistics.

Thus, based on the verification results of Eq.(26) we can make a conclusion, which generalizes the physical conditions of existence of Feoktistov's wave mode: the velocity of soliton-like wave propagation in neutron-multiplicating mediumin must be determined in general case by two conditions. The first condition (necessary) is predetermined by relationship between the equilibrium concentration and critical concentration of active (fissile) isotope $(\tilde{n}_{pu}/n_{crit}) > 1$ or, more exactly, by the Bohr-Sommerfeld quantization condition. The second condition (sufficient) is set by statistics of the Gaussian simplectic ensembles with respect to the parameter $a$, which describes the burning concentration wave width of active (fissile) component of nuclear fuel.

## V. COMPUTATION 3D-EXPERIMENT AND VERIFICATION OF THE WIGNER QUANTUM STATISTICS

Let us consider the simplified diffusion model of neutrons and nuclei kinetics in the chain (1) in the one-group approximation (neutron energy is ~ 1 MeV) and cylindrical geometry. Then, taking into account delayed neutrons, the respective system of differential equations, which describes the kinetics of Feoktistov's U-Pu fuel cycle, i.e., the kinetics of initiation and propagation of neutron-fission wave $n(x, t)$, is as follows [13]:

$$\frac{\partial n(x,t)}{\partial t} = D\Delta n(x,t) + q(x,t), \tag{27}$$

where

$$q(x,t) = [\nu(1-p)-1] \cdot n(x,t) \cdot \upsilon_n \cdot \sigma_f^{Pu} \cdot N_{Pu}(x,t) + \sum_{i=1}^{6} \frac{\tilde{N}_i \ln 2}{T_{1/2}^i} -$$

$$- n(x,t) \cdot \upsilon_n \cdot \left[ \sum_{8,9,Pu} \sigma_a^i \cdot N_i(x,t) + \sum_{i=1}^{6} \sigma_a^i \cdot \tilde{N}_i(x,t) + \sum_{i=осколки} \sigma_a^i \cdot \overline{N}_i(x,t) \right],$$

$$\frac{\partial N_8(x,t)}{\partial t} = -\upsilon_n \cdot n(x,t) \cdot \sigma_a^8 \cdot N_8(x,t), \tag{28}$$

$$\frac{\partial N_9(x,t)}{\partial t} = \upsilon_n \cdot n(x,t) \cdot \sigma_a^8 \cdot N_8(x,t) - \frac{1}{\tau_\beta} N_9(x,t), \tag{29}$$

$$\frac{\partial N_{Pu}(x,t)}{\partial t} = \frac{1}{\tau_\beta} N_9(x,t) - \upsilon_n \cdot n(x,t) \cdot (\sigma_a^{Pu} + \sigma_f^{Pu}) \cdot N_{Pu}(x,t), \tag{30}$$

$$\frac{\partial \tilde{N}_i}{\partial t} = p_i \cdot \upsilon_n \cdot n(x,t) \cdot \sigma_f^{Pu} \cdot N_{Pu}(x,t) - \frac{\ln 2 \cdot \tilde{N}_i}{T_{1/2}^i}, \quad i=1,6. \tag{31}$$

To determine the last term $q(x, t)$ on the right-hand-side of Eq.(27), we use the effective additional neutron absorber approximation:

$$n(x,t) \cdot \upsilon_n \cdot \sum_{i=осколки} \sigma_a^i \cdot \overline{N}_i(x,t) = n(x,t) \cdot \upsilon_n \cdot \sigma_a^{eff} \cdot \overline{N}(x,t). \tag{32}$$

Taking into account the fact that fission with two fragment formation is most probable, the kinetic equation for $\overline{N}(x,t)$ becomes

$$\frac{\partial \overline{N}(x,t)}{\partial t} = 2\left(1 - \sum_{i=1}^{6} p_i\right) \cdot n(x,t) \cdot \upsilon_n \cdot \sigma_f^{Pu} \cdot N_{Pu}(x,t) + \sum_{i=1}^{6} \frac{\tilde{N}_i \ln 2}{T_{1/2}^i}. \tag{33}$$

Here $n(x,t)$ is the neutron density; $D$ is the diffusion constant of neutrons; $\upsilon_n$ is the neutron velocity ($E_n = 1$ MeV in the one-group approximation); $\tilde{N}_i$ are the concentrations of neutron-rich fission fragments of the $^{239}Pu$ nuclei; $N_8$, $N_9$, $N_{Pu}$ are the $^{238}U$, $^{239}U$, $^{239}Pu$ concentrations; $\overline{N}_i$ are the concentrations of rest fission fragments of the $^{239}Pu$ nuclei; $\sigma_a$ is the neutron-capture microcross-section; $\sigma_f$ is the fission microcross-section; $\tau_\beta$ is the nucleus life time with respect to the $\beta$–decay; $p_i$ ($p = \sum_{i=1}^{6} p_i$) are the parameters characterizing delayed neutrons groups for main fuel fissionable nuclides [18].

The boundary conditions for the system of differential equations (27)-(31) are

$$n(x,t)\big|_{x=0} = \Phi_0/\upsilon_n, \quad n(x,t)\big|_{x=l} = 0, \tag{34}$$

where $\Phi_0$ is the neutron density of plane diffusion source of neutrons which is located on the boundary $x=0$; $l$ is the uranium block length.

An estimation of the neutron flux density $\Phi_0$ from the external source on the boundary can be obtained from an estimation of the Pu critical concentration which is of order of 10%:

$$4\tau_\beta \Phi_0 \sigma_a^8 N_8(x,t)\big|_{t=0} = 0.1 N_8(x,t)\big|_{t=0}, \tag{35}$$

and therefore

$$\Phi_0 \approx 0.1/4\tau_\beta \sigma_a^8. \tag{36}$$

Here we note that Eq. (36) is only an estimation of $\Phi_0$. The results of computational experiment show that it can be substantially smaller in reality.

In general, different boundary conditions can be used, depending on physical conditions under which nuclear burning is initiated by the source neutrons, for example, the Dirichlet condition of (36) type, the Neumann condition or the so-called third-kind boundary condition, which summarizes the first two conditions. Use of the third-kind boundary condition is recommended in neutron transport theory [18]. Here we use this condition in the simple case which is known as Milne's problem, or more precisely, it is the linear combination of the neutron concentration $n(x,t)$ and its spatial derivative $\partial n/\partial x(x,t)$ on the boundary:

$$n(0,t) - 0.7104\lambda n^{(1,0)}(0,t) = 0, \tag{37}$$

where $\lambda$ is the range of neutrons and $n^{(1,0)}(0, t) \equiv \partial n/\partial x (0, t)$.

Although the behavior of the "neutron source-nuclear fuel" system depends on the boundary conditions near the boundary, computational experiments show that in reactor core, i.e., far from the boundary, the system is asymptotically independent of the boundary conditions. This confirms the independence of wave propagation in reactor volume on the boundary conditions and parameters of nuclear fuel "ignition". In this sense the problem of determining the optimum parameters of nuclear fuel "ignition" in "neutron source-nuclear fuel" system is a nontrivial and extraordinarily vital issue, which requires a separate examination.

The initial conditions for the system of differential equations (27)-(31) are

$$n(x,t)\big|_{x,t=0} = \Phi_0/\upsilon_n, \quad n(x,t)\big|_{x,t=0} = 0; \tag{38}$$

$$N_8(x,t)\big|_{t=0} = \frac{\rho_8}{\mu_8}N_A \approx \frac{19}{238}N_A, \tag{39}$$

$$N_9(x,t)\big|_{t=0} = 0, \quad N_{Pu}(x,t)\big|_{t=0} = 0, \quad \tilde{N}_i(x,t)\big|_{t=0} = 0, \quad \overline{N}(x,t)\big|_{t=0} = 0, \tag{40}$$

where $\rho_8$ is the density, which is expressed in the units of g·cm$^{-3}$; $N_A$ is the Avogadro constant.

The following values of constants were used for simulation:

$$\sigma_f^{Pu} = 2{,}0\cdot 10^{-24}\ cm^2;\ \sigma_f^8 = {,}55\cdot 10^{-24}\ cm^2; \tag{41}$$

$$\sigma_a^8 = \sigma_a^i = \sigma_a^{fragments} = 5.38\cdot 10^{-26}\ cm^2;\ \sigma_a^9 = \sigma_a^{Pu} = 2.12\cdot 10^{-26}\ cm^2; \tag{42}$$

$$\nu = 2{,}9;\ \tau_\beta \sim 3{,}3\ \text{days};\ \upsilon_n \approx 10^9\ \text{cm/s};\ D \approx 2.8\cdot 10^9\ \text{cm}^2/\text{s}. \tag{43}$$

The system of equations (27)-(32) with boundary conditions (37)-(35), initial conditions (38)-(40) and the values of constants (41)-(43) is solved numerically using the software package Fortran Power Station 4.0. At the same time we use the DMOLCH subprogram from the IMSL Fortran Library. The DMOLCH subprogram solves a system of partial differential equations of the form $u_t=f(x,t,u_x,u_{xx})$ by the method of straight lines [13, 19]. The solutions of diffusion model of neutrons and nuclei kinetics in the chain (1) in the one-group approximation and cylindrical geomerty are presented in Fig.5.

Verification of the Wigner symplectic statistics consists in comparison of the experimental velocity of nuclear burning wave obtained by a computational 3D-experiment with its theoretical value obtained by Eq. (26). For this purpose we at first find the plutonium critical concentration $n_{crit}^{fis}$ from the profile of space-time evolution of its experimental concentration distribution (Fig. 5). It is obvious, that the absolute value of critical concentration approximately is $N_{crit}^{Pu} \cong 8\cdot 10^{20}$ cm$^{-3}$·(see Fig. 6(b)). It follows that the plutonium normalized critical concentration is

$$n_{crit}^{fis} = N_{crit}^{Pu}/N_8(x,0) = 0.0167, \tag{44}$$

where by virtue of Eq.(39) the initial uranium concentration is $N_8(x,0)= 4.79 \cdot 10^{22}$ cm$^{-3}$ and the value of $a_*$ is equal to 0.704 by virtue of Eq. (25). In other words, the important case when $a_*<1$ takes a place (see Fig. 4).

Taking into account the plutonium normalized equilibrium concentration $\tilde{n}_{fis} = 0.1$, by virtue of Eq. (26) we have the theoretical value of the Wigner symplectic probability:

$$\frac{1}{2}\Lambda(a_*) = p_W^s(a_*) = 0.9303, \quad (45)$$

which corresponds to the velocity of nuclear burning wave of $u_{theor} = 2.82$ cm/day at known parameters $L=5$ cm and $\tau_\beta = 3.3$ days.

Now we can simply determine the experimental values of nuclear burning wave velocity and, accordingly, the Wigner symplectic probability. In Fig. 6(a) the profile of space-time evolution of experimental concentration distribution of neutrons is shown. We can see that the wave crest has covered the distance of 600 cm during $t=217$ days. So, the velocity of nuclear burning neutron wave is

$$u_{simul} = 600/217 \cong 2.77 \quad cm/day. \quad (46)$$

This, in its turn, corresponds to the value of $(1/2)\Lambda(a_*) = p_W^s(a_*) = 0.9141$.

Thus, the approximate equality of the experimental and theoretical velocity of nuclear burning wave ($u_{theor} \cong u_{simul}$) makes it possible to conclude that the Wigner quantum (symplectic) statistics verified by computing 3D-experiment (see Fig. 4) satisfactorily describes experimental data characterized by the parameter $\Lambda(a_*)$.

Here we note that computing experiments show that the conditions of wave blocking, which describe the degradation and subsequent stop of wave, are predetermined by the degree of burn-up of the main nonfissionable ($^{238}$U) and fissionable ($^{239}$Pu) components of nuclear fuel in front of the wave by neutrons from external source in the initial stage of wave "ignition". This process is very important, since the high degree of fuel component burn-up in front of the wave will inhibit the wave from overcoming this region just as fire in the steppe can not cross the plowed in advance stripe of the land. It is obvious that in the initial stage of wave initiation the degree of fuel burn-up is determined first of all by the energy spectrum and intensity of neutrons from the external source and by the properties of nuclear fuel. The most important of these properties is the delay time $\tau_\beta$ of active (fissile) isotope generation due to the $\beta$–decay of compound nuclei in the Feoktistov U-Pu fuel cycle (1) or the Teller Th-U fuel cycle (2).

In spite of the general understanding of physics of nuclear burning wave blocking, it is obvious that indicated above difficulties in the describing this process testify to nontriviality of given problem. Unfortunately, the solving of this problem exceeds the scope of this work, but it will be a subject of future research.

## CONCLUSIONS

The solutions of the system of diffusion type equations for neutrons and concomitant kinetic equations for nuclei obtained by numerical 3D-simulation persistently point to the regions where the stable soliton-like solutions for neutrons and solitary wave solutions for nuclei are existed. This is no wonder for nearly intergrable systems, to which the investigated system of equations for neutrons and nuclei belongs, whereas the existence of stable soliton-like solutions in three spatial dimensions causes a surprise for the following reason.

As is known, the derivation and solution of integrable nonlinear evolution partial differential equations in three spatial dimensions has been the holy grail in the field of integrability since the late 1970s. The celebrated Korteveg-de Vries and nonlinear Schrödinger equations, as well as Kadomtsev-Petviashvili and Davey-Stewertson equations, are prototypical examples of integrable evolution equations in the one and two spatial dimensions, respectively. Do there exist integrable analogs of these equations in three spatial dimensions?

As it has turned out, quite recently, in 2006, the method for finding of an analytical solutions of indicated above partial differential equations in three spatial dimensions was developed [20]. Therefore, the natural question arises: "To which from this equations does the diffusion equation for neutrons correspond, or, maybe, his is perfectly a new type of soliton partial differential equations in three spatial dimensions?"

## ACKNOWLEDGMENT

The authors acknowledge the financial support of the DFG through SFB 701 "Spectral structures and topological methods in mathematics", Bielefeld University (Germany), German-Ukrainian Project 436 UKR 113/94, and IGK Bielefeld (Germany).

TABLE I. The parameters of nuclear burning wave

| Parameter | U-Pu cycle | | | | | | | Th-U cycle |
|---|---|---|---|---|---|---|---|---|
| | References | | | | | | | |
| | Present paper | [14] | [15] | [15] | [16] | [4] | [17] | *) |
| $\dfrac{\tilde{n}_{equil}^{fis}}{n_{crit}^{fis}}$ | $\dfrac{0.100}{0.017}$ | $\dfrac{2.585}{1.750}$ | $\dfrac{0.145}{0.080}$ | $\dfrac{0.024}{0.015}$ | $\dfrac{0.240}{0.105}$ | $\dfrac{0.10}{0.05}$ | $\dfrac{0.071}{0.032}$ | $\dfrac{0.070}{0.035}$ |
| $a_*$ | 0.704 | 2.274 | 1.743 | 2.028 | 1.385 | 1.571 | 1.423 | 1.571 |
| $u_{theor}/u_{simul}$ [cm/year] | 1030/1012 | 2.9/3.1 | 125/130 | 21/22 | 622/620 | 293/331 | 46/~50 | 25 |

* Forecast for the Th−U fuel cycle in infinite medium at 10% enrichment of $^{233}$U.

# FIGURE CAPTIONS

FIG. 1. Time dependence of neutron concentration. Propagating wave (a) and locked wave (b): a segment of the curve of $n_{Pu}(z)$ above the $n_{cr}$ line is the reactor core; the scales of $n_{cr}$ and $n_{Pu}$ are given with $a \times 10$ magnification [4].

FIG. 2. The schematic view of permitted and subbarier (gray colored) region corresponding to the conditions $n_{Pu} > n_{crit}$ and $n_{Pu} < n_{crit}$, respectively. The delineated by square region is considered more particularly in Fig. 3.

FIG. 3. Schematic description of the permitted and forbidden region boundaries of nuclear burning according to the Borh-Sommerfeld condition (a) and the corresponding quasi-equivalent two-level scheme (b).

FIG. 4. The theoretical (solid line)) and experimental (points) dependence of $\Lambda(a_*)$ on the parameter $a_*$.

FIG. 5. Concentration kinetics of neutrons, $^{238}U$, $^{239}U$ and $^{239}Pu$ in the core of cylindrical reactor with radius of 125 cm and 1000 cm long at the time of 240 days. Here $r$ is transverse spatial coordinate axis (cylinder radius), $z$ is longitudinal spatial coordinate axis (cylinder length).

FIG. 6. (a) - The neutron concentration distribution at the cylinder axis at $t = 217$ days. The wave velocity is $u_{simul} \approx 2{,}77$. (b) - The $^{239}Pu$ concentration distribution at the cylinder axis for $n_{Pu} = 0.1 n_{crit}^{Pu}$ $n_{Pu} = 0.0167$ at $t = 217$ days.

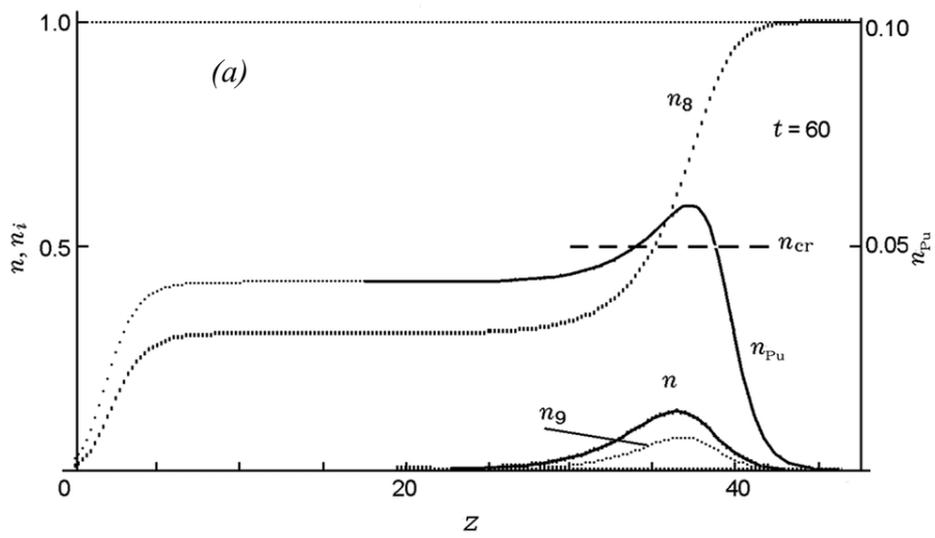

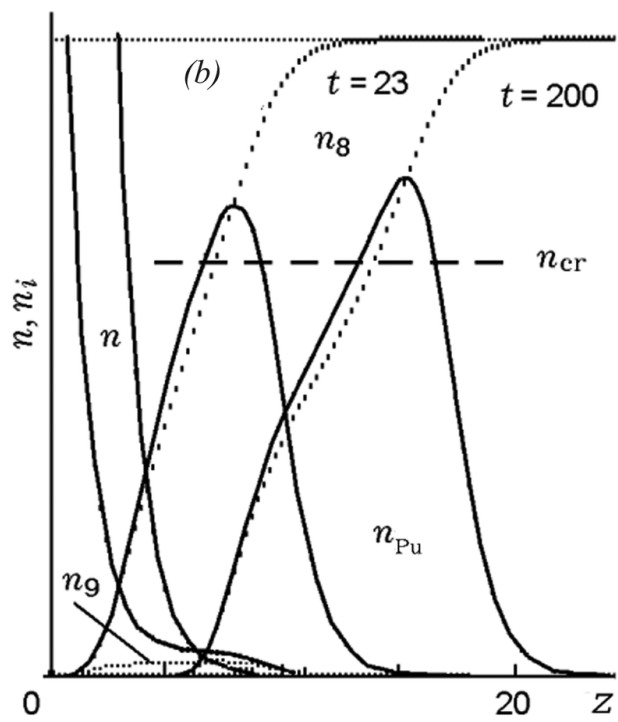

FIG. 1.

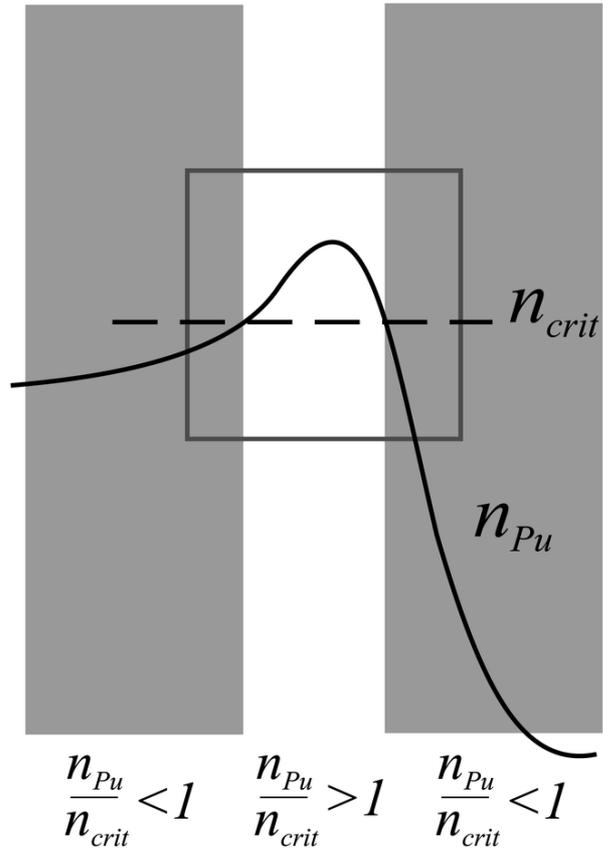

FIG. 2

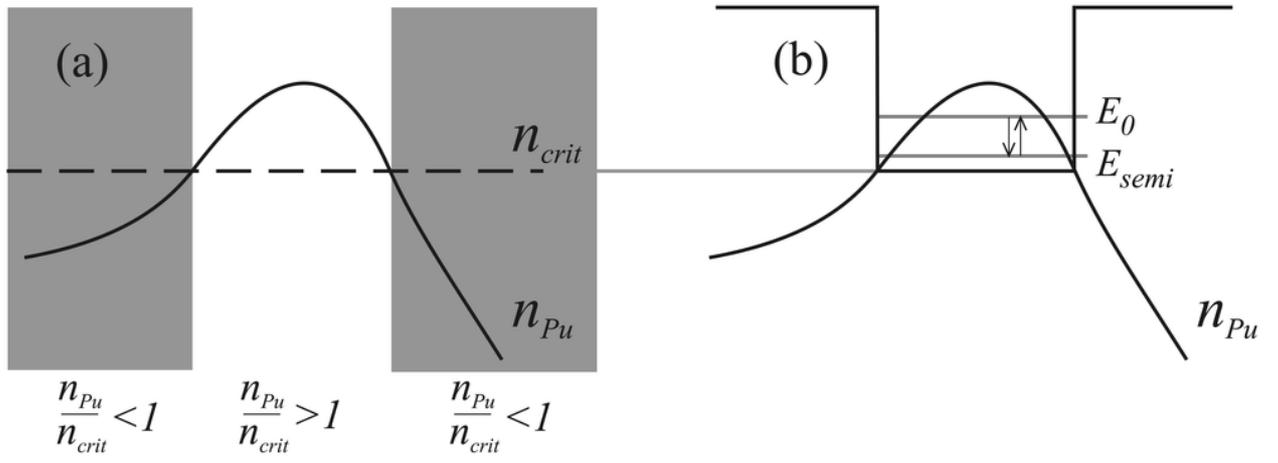

FIG. 3

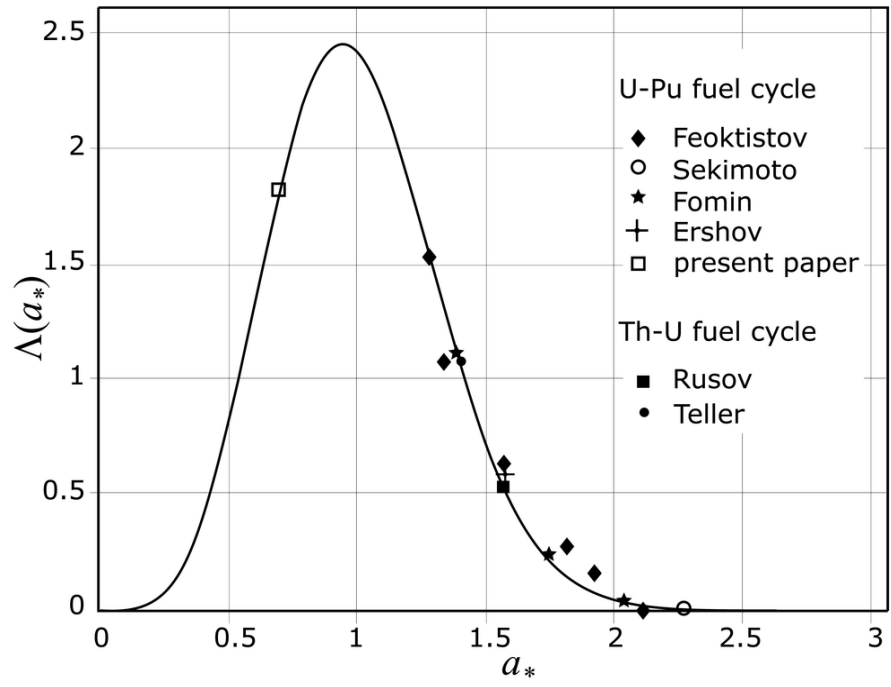

FIG. 4

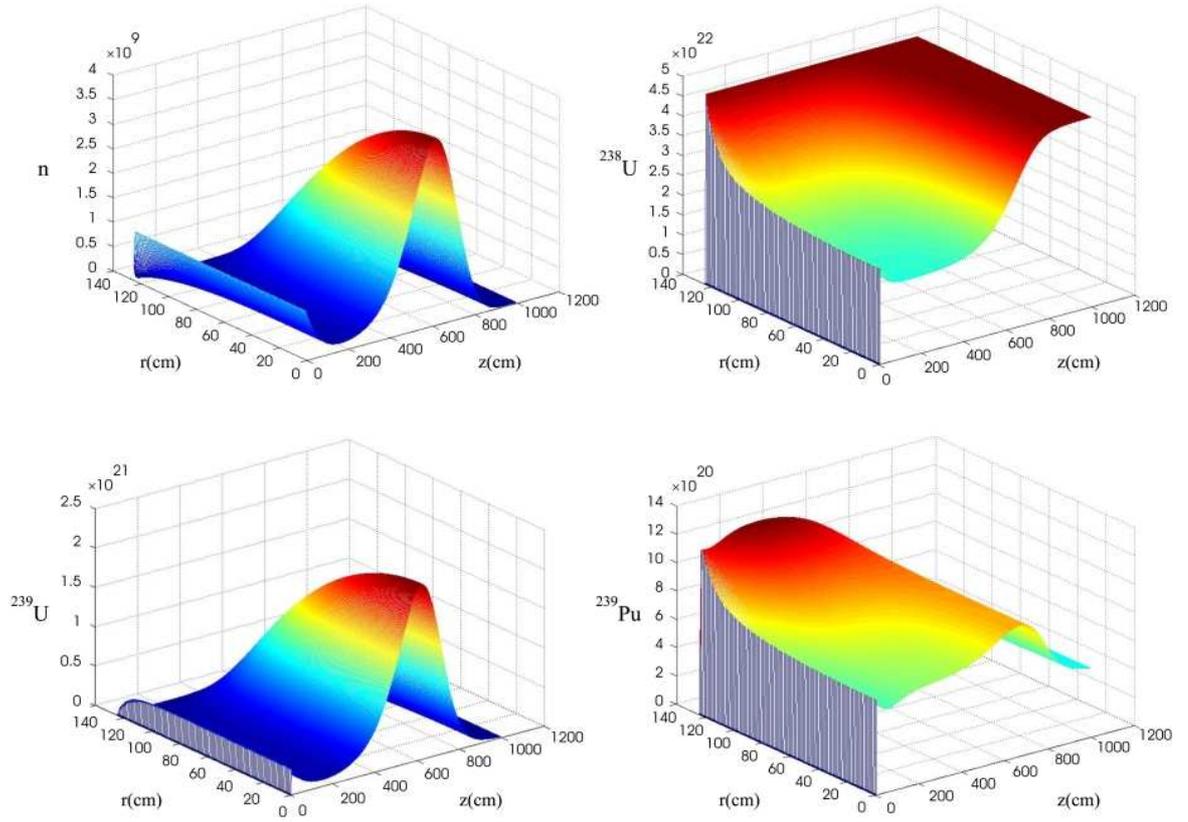

FIG. 5

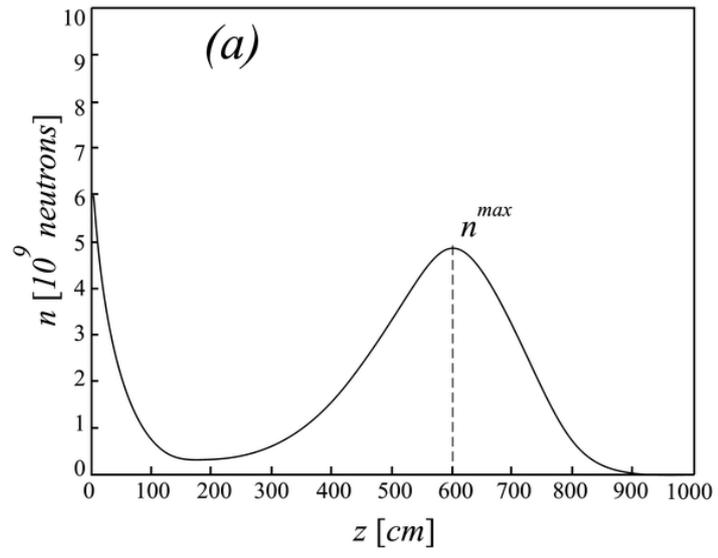

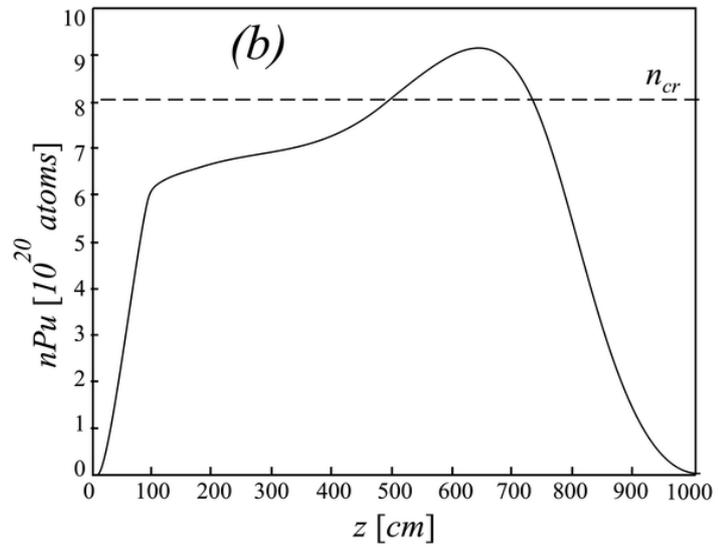

FIG. 6